\documentclass{iaus}
\usepackage{graphicx}

\title[Identifying Long-Period Transiting Planets with {\it Kepler}]
{Getting More For Your Money: Identifying and Confirming Long-Period Planets with {\it Kepler}}

\author[Jennifer C. Yee \& B. Scott Gaudi]
{Jennifer C. Yee \and B. Scott Gaudi}

\affiliation{Dept. of Astronomy, The Ohio State University \\ 140 W. 18th Ave., Columbus, OH, 43212, USA \\ email: {\tt jyee, gaudi@astronomy.ohio-state.edu}}

\pubyear{2008}
\volume{253}
\pagerange{000--000}
\jname{Transiting Planets}
\editors{F. Pont, J. Bland-Hawthorn \& B. Nordstrom, eds.}
\begin{document}

\maketitle

\begin{abstract}
{\it Kepler} will monitor enough stars that it is likely to detect single transits of planets with periods longer than the mission lifetime. We show that by combining the {\it Kepler} photometry of such transits with precise radial velocity (RV) observations taken over $\sim$3 months, and assuming circular orbits, it is possible to estimate the periods of these transiting planets to better than 20\% (for planets with radii greater than that of Neptune) and the masses to within a factor of 2 (for planet masses $m_{\rm p} \ge M_{\rm Jup}$). We also explore the effects of eccentricity on our estimates of these uncertainties.
 
\keywords{methods: analytical, planetary systems, planets and satellites: general}

\end{abstract}

\firstsection

\section{Introduction}
Planets that transit their stars offer us the opportunity to study the
physics of planetary atmospheres and interiors, which may help
constrain theories of planet formation. From the photometric light
curve, we can measure the planetary radius and also the orbital
inclination, which when combined with radial velocity (RV) observations,
allows us to measure the mass and density of the planet. All of the known transiting planets orbit so close to their parent stars that the stellar flux plays
a major role in heating these planets. In contrast, the detection of
transiting planets with longer periods ($P > 1\, {\rm yr}$) and
consequently lower equilibrium temperatures would allow us to probe a
completely different regime of stellar insolation, one more like that
of Jupiter and Saturn whose energy budgets are dominated by their
internal heat.

Because of its long mission lifetime ($L = 3.5\, {\rm yrs}$),
continuous observations, and large number of target stars ($N \simeq
10^5$), the {\it Kepler} satellite (Borucki et al. 2004; Basri et al. 2005) has a unique opportunity to
discover long-period transiting systems. Not only will {\it Kepler}
observe multiple transits of planets with periods up to the mission
lifetime, but it is also likely to observe single transits of planets
with periods longer than the mission lifetime. As the period of a
system increases beyond $L/2$, the probability of observing more than
one transit decreases until, for periods longer than $L$, only one transit
will ever be observed.  For periods longer than $L$, the probability
of seeing a single transit diminishes as $P^{-5/3}$.  Even with only a
single transit observation from {\it Kepler}, these long-period
planets are invaluable.

\begin{figure}
  \begin{center}
  \includegraphics[width=4in]{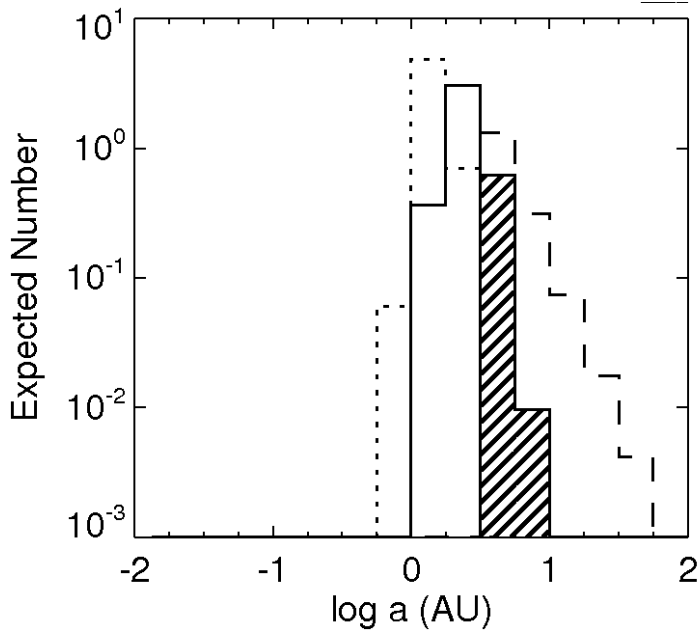}
  \end{center}
  \caption{The expected number of one (solid line) and two (dotted line) transit events. The probability of observing exactly one or exactly two transits during the mission lifetime is convolved with the known distribution of semi-major axes. The shaded area indicates where this distribution is expected to be incomplete, and the dashed line shows our extrapolation to larger semi-major axes based on the flat distribution fit by Cumming et al. (2008). \label{f1}}
\end{figure}

Figure \ref{f1} gives the total number of systems that {\it Kepler} should observe that exhibit exactly one and exactly two transits during the mission lifetime of 3.5 years. The probability of observing a transit during the mission has been convolved with the observed fraction of planets as a function of semimajor axis for $a \lesssim 3\, {\rm AU}$ and multiplied by the number of stars {\it Kepler} will observe (100,000) to give the number of single transit systems and two transit systems that can be expected. Above $a \gtrsim 3\, {\rm AU}$ the observed sample of planets as a function of semi-major axis is expected to be incomplete. We have extrapolated the number of expected single transit systems to larger semi-major axes by convolving the single transit probability distribution with the extrapolation of the fraction of planets as a function of semi-major axis by Cumming et al. (2008) that is constant in dN/dloga. The total expected numbers are 4.0 one transit events (5.7 with the extrapolation) and 5.6 two transit events. 

\begin{figure}
  \begin{center}
    \includegraphics[width=4in]{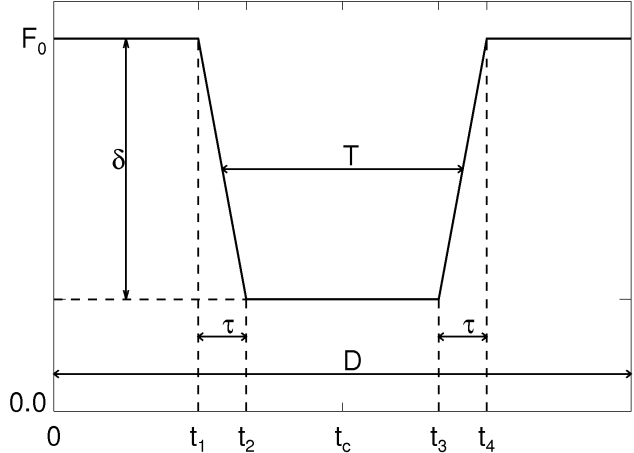}
  \end{center}
  \caption{The model lightcurve. $t_c$ is the time of transit center, $T$ is the duration of the full-width half-maximum depth of the transit, $\tau$ is duration of ingress or egress, and $\delta$ is the fractional depth of the transit. $t_1$--$t_4$ are the points of contact. $F_0 \equiv 1$ is the out-of-transit flux, and $D$ is the total duration of the observations. \label{f2}}
\end{figure}

\section{Overview}
We derive exact expressions for the uncertainties in $t_c$, $T$, $\tau$,
and $\delta$,  by applying the Fisher matrix formalism to the simple light curve model shown in Fig. \ref{f2} (Gould 2003). For the long-period transiting planets observed by {\it Kepler}, $ \tau \ll T \ll D$ (the total duration of the observations), so the largest uncertainty is for $\tau$:
\begin{equation}
\frac{\sigma_{\tau}}{\tau} = Q^{-1}\sqrt{\frac{6T}{\tau}} .
\end{equation}
Here $Q$ is approximately equal to the total signal-to-noise ratio of the transit,
\begin{equation}
Q \equiv (\Gamma_{\rm ph} T)^{1/2} \delta,
\end{equation}
where $\Gamma_{\rm ph}$ is the photon collection rate. For a Jupiter-sized planet with period equal to the mission lifetime and transiting a $V=12$ solar-type star, $Q \simeq 1300$, whereas for a Neptune-sized planet, $Q \simeq 150$.

Assuming the planet is in a circular orbit, and the stellar density, $\rho_{\star}$, is known from spectroscopy, Seager \& Mall\'{e}n-Ornelas (2003) showed that the planet's period, $P$, can be derived from a single observed transit,
\begin{equation}
P = \frac{2\pi a}{v_{\rm tr,p}} = \frac{G\pi^2}{3}\rho_{\star}\left(\frac{T\tau}{\sqrt{\delta}}\right)^{3/2}
\end{equation}
where $v_{\rm tr,p}$ is the velocity of the planet at the time of transit.

\begin{figure}
\includegraphics[width=2.65in]{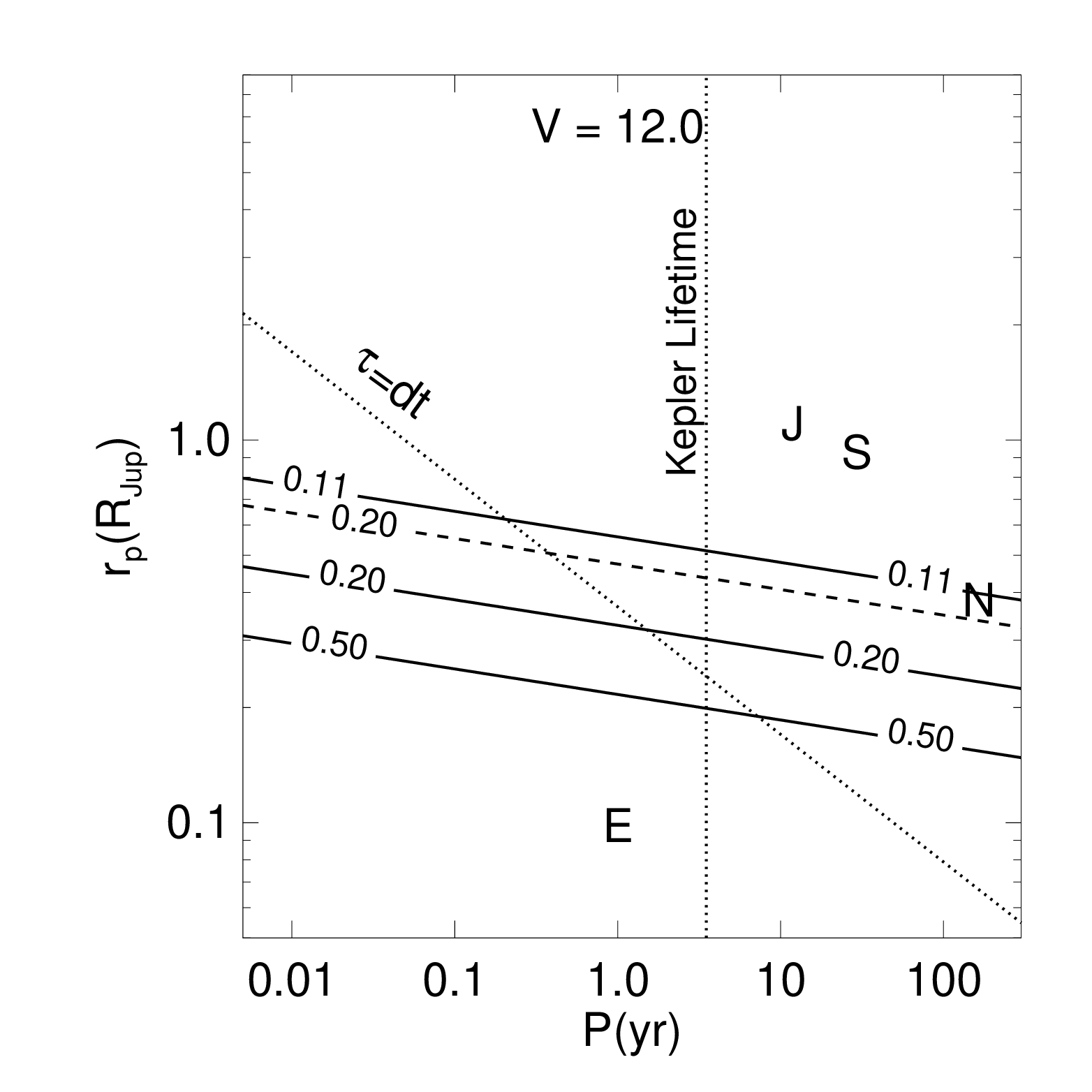}\includegraphics[width=2.65in]{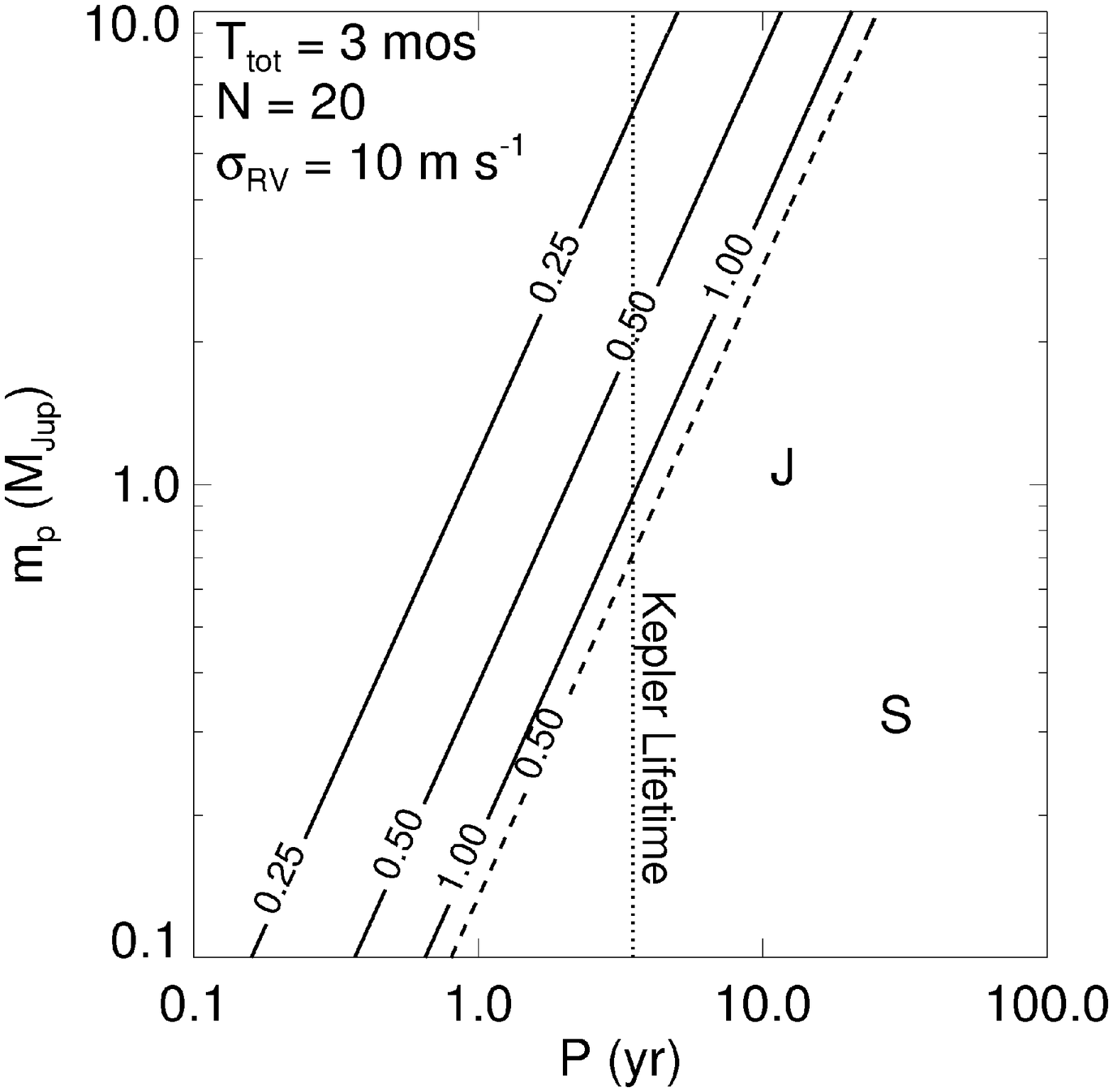}
\caption{{\bf Left}: Contours of constant fractional uncertainty in $P$: $\sigma_P/P$. {\bf Right}: Contours of constant fractional uncertainty in $m_{\rm p}$: $\sigma_{m_{\rm p}}/m_{\rm p}$. \label{f6}}
\end{figure}

Figure \ref{f6} shows contours of constant fractional uncertainty in the period as a function of $P$ and the planet radius, $r_{\rm p}$. The model assumes a solar-type star with $V=12.0$, an impact parameter b=0.2, and a 10\% uncertainty in $\rho_{\star}$. The dashed line indicates how the fractional uncertainty in $P$ changes with $V$; it shows the 0.20 contour for $V=14.0$ and $b=0.2$. The diagonal dotted line shows the boundary below which the assumptions of the Fisher matrix break down (the ingress/egress time $\tau$ is roughly equal to the sampling rate of $1/dt =$ 1 observation/30 mins). The vertical dotted line shows the mission lifetime of {\it Kepler} ($L=$3.5 yrs). Solar system planets are indicated. 

The mass of the planet can be estimated by sampling the stellar radial velocity (RV) curve soon after the transit is observed. Near the time of transit we can expand the stellar RV

\begin{equation}
v_{\star} \approx v_0 - A_{\star}(t-t_c),
\end{equation}
where $A_{\star}=2\pi K_{\star}/P$, and 
\begin{equation}
K_{\star} = \left(\frac{2 \pi G}{PM_{\star}^2}\right)^{1/3} m_{\rm p} \sin i
\end{equation}
is the stellar radial velocity semi-amplitude and we assume $\sin i \approx 1$.
The mass of the planet, $m_{\rm p}$, can be determined from the observables, $g_{\star}$, $A_{\star}$, $T$, $\tau$, and $\delta$:

\begin{equation}
m_{\rm p} = \frac{1}{16G}g_{\star}^2A_{\star}\left(\frac{T\tau}{\sqrt{\delta}}\right)^2.
\end{equation}

Figure \ref{f6} shows contours of constant uncertainty in $m_{\rm p}$ as a function of $m_{\rm p}$ and $P$. The model assumes a 10\% uncertainty in $g_{\star}$. The solid lines show the result for 20 radial velocity measurements with precision of 10 m s$^{-1}$ taken over a period of 3 months after the transit. The dashed line shows the 0.50 contour for 40 observations taken over 6 months. The dotted line indicates the mission lifetime of 3.5 years. The positions of Jupiter and Saturn are indicated.

If the eccentricity $e$ is non-zero, there is not a unique solution for $P$ and $m_{\rm p}$, however, we can still constrain them by adopting priors on $e$ and the argument of periastron $\omega_{\rm p}$. Using $e=0.3$ gives a range of inferred periods of 0.4--2.5 relative to the assumption of a circular orbit. The range of inferred masses is 0.5--1.7, which is {\it smaller} than the contribution from the uncertainty in $A_{\star}$. 

In this paper, we demonstrated that it will be possible to detect and characterize long-period planets using observations of single transits by the {\it Kepler} satellite, combined with precise radial velocity measurements taken immediately after the transit. These results can also be generalized to any transiting planet survey, and it may be particularly interesting to apply them to the {\it COnvection, ROtation \& planetary Transits (CoRoT)} mission (Baglin 2003).  Additionally, estimates of the period of a transiting planet could allow the {\it Kepler} mission to make targeted increases in the sampling rate at the time of a second transit occurring during the mission.

\end{document}